\documentstyle[aps,twocolumn]{revtex}

\newcommand{\bx}{\mbox{\boldmath $x$}}
\newcommand{\blk}{\mbox{\boldmath $k$}}
\newcommand{\ssum}{\sum_n{'}}
\newcommand{\k}{k}

\begin{document}

\title{Evolution of a Bose-Einstein condensate of neutral atoms\\
--- A field theoretical approach}

\author{Koichi NAKAMURA}

\address{Division of Natural Science, Meiji University, Izumi Campus\\
Eifuku, Suginami-ku,Tokyo 168-8555, JAPAN\\
knakam@isc.meiji.ac.jp}

\author{Yoshiya YAMANAKA}

\address{Waseda University Senior High School\\
Kamishakujii, Nerima-ku, Tokyo 177-0044, JAPAN\\
yamanaka@mn.waseda.ac.jp}

\maketitle

\begin{abstract}                

The particle distribution in a Bose condensate
under the trapping potential and its time evolution   
after switching off the trapping potential suddenly are calculated. We investigate
the problem from the viewpoint of quantum field theory,@using 
a model of self-interacting 
neutral boson field.  Within the approximation of retaining the most 
dominant term in the Hamiltonian after applying 
the Bogoliubov replacement, we can calculate analytically the 
particle distribution as a function of space and time coordinates.

\end{abstract}

\pacs{ }

The recent observations of Bose-Einstein
condensation (BEC) in magnetically 
trapped atomic gases \cite{JILA,MIT} has stimulated the theoretical 
study of this phenomena.  In this work we investigate the
particle distribution in a condensate of neutral boson field
under the trapping potential and its time evolution immediately  
after the trapping potential is switched off, 
taking the interaction between the particles into account.

We consider a collection of interacting Bose particles  
trapped in a harmonic oscillator potential $\frac12 \sum_{i=1}^3 
\omega_i^2 x_i^2$, whose Hamiltonian is given by
\begin{eqnarray}
H_B\!\!&=&\!\!\int\! d\bx\, \phi^\dagger(\bx)\left\{-\frac12\triangle+
      \frac12\sum_{i=1}^{3}\omega_i^2x_i^2\right\}\phi(\bx)\nonumber\\
   & &\!\!+\frac12 g\!\int\! d\bx d\bx'\, 
   \phi^\dagger(\bx)\phi^\dagger(\bx')V(\bx-\bx')
      \phi(\bx)\phi(\bx'),
\end{eqnarray} 
where we set particle mass$=\hbar=1$.

In a classical paper on BEC \cite{Bog}, Bogoliubov treated 
a similar Hamiltonian but 
without the trapping potential. The key idea in his theory is that the 
creation and annihilation operators for the zero momentum  
particle  should be
replaced by a $c$-number $\sqrt{N_0}$, when BEC occurs. The quantity 
$N_0$ is the average number of particles 
occupying the zero momentum state and 
BEC means that this number becomes very large and macroscopic.
Roughly speaking, this 
replacement is justified as follows:
For BEC state, $N_0$ particles are in the 
zero momentum state. Thus the creation and annihilation operators for zero 
momentum state give
\begin{eqnarray}
c_0^\dagger \vert N_0 > &=&\sqrt{N_0+1}\vert N_0+1> \nonumber \\ 
c_0 \vert N_0>&=&\sqrt{N_0}\vert N_0-1>.
\end{eqnarray}
Since the number $N_0$ is very large, $N_0+1\approx N_0$. Therefore
both the operations of $c_0^\dagger$ and $c_0$ on system states
are similar, giving the result $\sqrt{N_0}$ as an eigenvalue. 
Thus we can treat both the operators $c_0^\dagger$ and $c_0$ as $c$-number 
$\sqrt{N_0}$. 

To extend this idea to a trapped Bose gas, first let us introduce
annihilation operators of particles $a_n$ in 
the harmonic oscillator modes by expanding the field 
operator $\phi(\bx)$ in terms of a complete set 
 of the harmonic oscillator wave functions
$u_n(\bx)$:
\begin{equation} 
\phi(\bx)=\sum_na_nu_n(\bx),
\end{equation}
where $u_n(\bx)$ satisfies
\begin{equation}
\{-\frac12\triangle+\frac12\sum_{i=1}^{3}\omega_i^2 x_i^2\}u_n(\bx)=\varepsilon_n u_n(\bx),
\end{equation}
and $n=(n_1,n_2,n_3)$ with non-negative integers $n_i$.

In a condensed state, most particles are in the ground state of 
the harmonic 
oscillator, that is, the state for $n=(0,0,0)$. Therefore, the occupation 
number $N_0$ of this state is supposed to be very large. In typical 
experiments, the total particle number is of order of $10^3 \sim 10^5$. 
Then we can apply the Bogoliubov replacement to the creation 
and annihilation 
operators, $a_0^\dagger$ and $a_0$ :
\begin{equation}
a_0^\dagger\to\sqrt{N_0},\ \ \ \ a_0\to\sqrt{N_0}.
\end{equation}
Then the field operator $\phi(\bx)$ reads
\begin{equation} 
\phi(\bx)=\sqrt{N_0}u_0(\bx)+\sum_n{'} a_n u_n(\bx),
\label{BogField}
\end{equation}
where the prime in $\ssum$ means to omit the term $n=(0,0,0)$. 

Substituting the above expression of $\phi(\bx)$ into $H_B$ in (1),
we have 
\begin{eqnarray}
H_B\!\!&=&\!\!N_0\varepsilon_0+\sum_n{'}\varepsilon_n a^\dagger_n a_n
\nonumber \\
&& +\frac12 g N_0^2 \!\int\! d \bx d \bx'\, u_0^2(\bx)V(\bx-\bx')
      u_0^2(\bx')\nonumber\\
&& + g N_0^{3/2} \sum_n{'}(a^\dagger_n+a_n)\nonumber \\ 
&&\,\,\, \times  \!\int\! d\bx d\bx'\, u_n(\bx)u_0(\bx)V(\bx-\bx')
                    u_0^2(\bx')\nonumber\\
& &+O(gN_0)
\label{approxH}  
\end{eqnarray}
when $V(\bx)=V(-\bx)$ is assumed.
Since $N_0$ is large, we shall retain only terms of order $N_0^2$ and 
$N_0^{3/2}$ in the interaction part. We remark the presence of the terms
of order $N_0^{3/2}$.  In the original Bogoliubov
theory, the terms of order $N_0^{3/2}$ do not exist, which is due to the
momentum conservation. In the present case, however, we do not have such a 
conservation law, and the dominant $q$-number terms are 
linear in the creation and annihilation 
operators of excited states. 

By defining operators $\alpha_n^\dagger$ and $\alpha_n$ by 
\begin{equation}
\alpha_n=a_n+\sqrt{N_0}\gamma_n,\ \ \ \ \ \alpha_n^\dagger=a_n^\dagger
         +\sqrt{N_0}\gamma_n
\end{equation}
which satisfy the same commutation relations as $a_n^\dagger$ and $a_n$,
we can diagonalize the approximated Hamiltonian thus obtained: 
\begin{equation}
H_B=\sum_n{'}\varepsilon_n\alpha_n^\dagger\alpha_n+N_0
\varepsilon_0(1+\frac12
     \gamma_0-\sum_n{'}\frac{\varepsilon_n}{\varepsilon_0}\gamma_n^2),
\label{HB1}
\end{equation}
with the $c$-number coefficient
\begin{equation}
\gamma_n\equiv\frac{gN_0}{\varepsilon_n}\!\!\!\int\! 
d\bx d\bx'\, u_n(\bx)u_0(\bx)V(\bx-\bx')u_0^2(\bx').
\label{gamman}
\end{equation}

In terms of these $\alpha$'s, the particle number density operator 
$\hat{n}(\bx)$ is written as 
\begin{eqnarray}
\hat{n}(\bx)\!\!&\equiv&\!\! \phi^\dagger(\bx)\phi(\bx)\nonumber\\
&=&\!\!\sum_{m,n}{'}\alpha_n^\dagger\alpha_mu_n(\bx)u_m(\bx)
                        \nonumber\\
             & &\!\!+\sqrt{N_0}\sum_n{'}(\alpha_n^\dagger+\alpha_n)u_n(\bx)
             \nonumber \\
&& \,\,\,\times \left\{u_0(\bx)-\sum_m{'}\gamma_mu_m(\bx)\right\}\nonumber\\
             & &\!\!+N_0\left(u_0(\bx)-\sum_n{'}\gamma_nu_n(\bx)
             \right)^2   
\end{eqnarray}
and its thermal average, 
\begin{equation}
n(\bx)=Z^{-1}\mbox{\bf Tr}\left[e^{-\beta H_B}\hat{n}(\bx)\right]
\end{equation}
with $Z=\mbox{\bf Tr}\left[e^{-\beta H_B}\right]$, can easily
be calculated 
by making use of (\ref{HB1}) to find 
\begin{eqnarray}
n(\bx)&=&N_0\left(u_0(\bx)-\sum_n{'}\gamma_nu_n(\bx)\right)^2
\nonumber \\
&& +\sum_n{'}\frac{1}{e^{\beta  \varepsilon_n}-1}\,u_n^2(\bx).
\label{n2}
\end{eqnarray}

In our approximation of taking only the linear terms in 
$a_n^\dagger$ and $a_n$, the energy spectrum of single particle 
states are not affected by the existence of interactions between 
particles but the wave function of the ground state {\em does} change. 
This can be seen in the first term above. 

The total particle number is also obtained by integrating the above 
$n(\bx)$ with respect to $\bx$: 
\begin{equation}
N=N_0(1+\ssum\gamma_n^2)+\ssum \frac{1}{e^{\beta\varepsilon_n}-1}.
\label{N2}
\end{equation}
This equation gives the relation between the 
temperature and  the condensate
fraction $f_0=N_0/N$. It shows that the interaction between 
particles lowers the fraction $f_0$. 

Now we shall investigate how $n(\bx)$ evolves, if the trapping 
potential is  removed. Let us switch off the trapping potential 
at $t=0$, thereafter the system develops under the Hamiltonian, 
\begin{eqnarray}
H\!\!&=&\!\!\int\! d\bx\, \phi^\dagger(\bx)\left\{
-\frac12\triangle\right\}\phi(\bx)\nonumber\\
   & &\!\!+\frac12 g \!\int\! d\bx d\bx'\,\phi^\dagger(\bx)\phi^\dagger(\bx')V(\bx-\bx')
      \phi(\bx)\phi(\bx').
\label{H}
\end{eqnarray}
For a while, however, the system must be still in BEC. Therefore we may 
be able to approximate the self-interaction term (the second term) in 
(\ref{H}) 
by one in (\ref{approxH}) 
(to neglect the terms of order $g N_0$). Hence we have 
\begin{eqnarray}
H\!\!&\simeq&\!\!\int\! d\bx \phi^\dagger(\bx)\{-\frac12\triangle\}
    \phi(\bx)\nonumber\\
 & &\!\!+\sqrt{N_0}\ssum(a_n^\dagger+a_n)\varepsilon_n\gamma_n
    +N_0\varepsilon_0(1+\frac12\gamma_0). 
\label{approxH2}
\end{eqnarray}
Here $N_0$ and hence $\gamma_n$ must become time dependent. In principle, 
this dependence will be determined self-consistently by the condition 
that the total particle number is conserved. In the following, however,
our considerations are limited to the case of time-independent $N_0$ and 
$\gamma_n$, because we are interested in the short time evolution after
switching off the trapping potential. 

The quantum number $n$ does not diagonalize
the kinetic term in (\ref{H}).  It is convenient to introduce the 
creation and annihilation operators in momentum eigenstates 
$c_{k}^\dagger$ and $c_{\k}$ in a usual way: 
\begin{equation}
\phi(\bx)=\frac{1}{(2\pi)^{3/2}}\int\!\! d\blk\, c_{\k}
e^{i\blk \cdot \bx}.
\label{phick}
\end{equation}
Inversely we get 
\begin{eqnarray}
c_{\k}&=&\frac{1}{(2\pi)^{3/2}}\int\!\! d\bx e^{-i\blk \cdot \bx} \phi(\bx)
\nonumber \\
&=&\sqrt{N_0}\tilde u_0(\blk)+\ssum a_n \tilde u_n(\blk),
\label{ckan}
\end{eqnarray}
where we use the Bogoliubov replaced field operator (\ref{BogField}). 

The operators $a_n$ and $c_{\k}$ are related as, 
\begin{equation}
a_n=\int d\bx\, u_n(\bx) \phi(\bx)= \int\!\! d\blk\,
 c_{\k} \tilde u_n^\ast (\blk),
\end{equation}
where $\tilde u$ means Fourier transform of $u$ defined by 
\begin{equation}
\tilde f (\blk)=\frac{1}{(2\pi)^{3/2}}\int\!\!d\bx\, f(\bx)e^{-i\blk \cdot \bx}.
\end{equation}

Putting (\ref{phick}) and (\ref{ckan}) 
into (\ref{approxH2}), we have an expression of $H$ written by 
$c_{\k}^\dagger$ and $c_{\k}$ as 
\begin{eqnarray}
H\!\!&=&\!\!\int\! d\blk\, \varepsilon_{\k} c_{\k}^\dagger c_{\k}
\nonumber \\
&&+\sqrt{N_0}\int\!\! d\blk\, 
    \left\{ c_{\k}^\dagger \ssum \tilde u_n(\blk)\varepsilon_n\gamma_n
      +\mbox{h.c.}\right\}
      \nonumber\\
 & &+N_0\varepsilon_0(1+\frac12\gamma_0)     
\end{eqnarray}
with $\varepsilon_{\k}=\blk^2/2$.

The Heisenberg operator for $t>0$, 
\begin{equation}
c_{\k}(t)=e^{itH}\,c_{\k}\,e^{-itH}
\end{equation}
satisfies the equation of motion
\begin{eqnarray}
i\dot c_{\k}(t)&=&e^{itH}\,[c_{\k},H]\,e^{-itH}\nonumber \\
& =&\varepsilon_{\k}c_{\k}(t)+\sqrt{N_0}\ssum \varepsilon_n \gamma_n                       \tilde u_n(\blk).
\end{eqnarray}
We can solve this equation with the condition $c_{\k}(0)=c_{\k}$ to get 
\begin{equation}
 c_{\k}(t)=c_{\k}e^{-i\varepsilon_{\k}t}+\frac{e^{-i\varepsilon_{\k}t}-1}                      {\varepsilon_{\k}}\sqrt{N_0}\ssum \varepsilon_n \gamma_n 
              \tilde u_n(\blk).
\end{equation}
Substituting (\ref{ckan}) into the first term above, we have 
\begin{eqnarray}
 c_{\k}(t)\!\!&=&\!\!\ssum a_n \tilde u_n(\blk)e^{-i\varepsilon_{\k}t}+\sqrt{N_0} 
               \tilde u_0 (\blk)e^{-i\varepsilon_{\k}t}\nonumber\\
            & & \!\!+\frac{e^{-i\varepsilon_{\k}t}-1}{\varepsilon_{\k}}\sqrt{N_0}
               \ssum \varepsilon_n \gamma_n 
               \tilde u_n(\blk).
\end{eqnarray}
The field operator in the Heisenberg picture is written as 
\begin{eqnarray}
\phi(\bx,t)\!\!&=&\!\!e^{itH}\, \phi(\bx) \, e^{-itH}=\frac{1}{(2\pi)^{3/2}}\!\int\!
              d\blk\, c_{\k}(t) e^{i\blk \cdot \bx}\nonumber\\
           &=&\!\!\ssum a_n u_n(\bx,t)+\sqrt{N_0}u_0(\bx,t)
           \nonumber \\
           &&-i\sqrt{N_0}
               \ssum \varepsilon_n \gamma_n 
               \!\int_0^t\! d\lambda\,u_n(\bx,\lambda), 
\label{phi2}
\end{eqnarray}
where we denote 
\begin{equation}
u_n(\bx,t)=\frac{1}{(2\pi)^{3/2}} \!\int\! d\blk \,\tilde u_n(\blk) 
           e^{i\blk \cdot \bx - i\varepsilon_{\k}t}.
\label{defun}
\end{equation}
In terms of $\alpha_n$'s, (\ref{phi2}) can be rewritten as
\begin{eqnarray}
\phi(\bx,t)\!\!&=&\!\!\ssum \alpha_n u_n(\bx,t)+\sqrt{N_0}u_n(\bx,t)
              \nonumber\\
           & &\!\!-\sqrt{N_0}\ssum \gamma_n u_n(\bx,t)
           \nonumber \\
           &&-i\sqrt{N_0}
              \ssum \varepsilon_n \gamma_n\!\int_0^t d\lambda\,
               u_n(\bx,\lambda)      
\end{eqnarray}
and the number density operator in the Heisenberg picture is given by
\begin{eqnarray}
\hat n(\bx,t)\!\!&=&\!\!\phi^\dagger(\bx,t) \phi(\bx,t)\nonumber\\
             &=&\!\!\sum_{n,m}{'} \alpha_n^\dagger \alpha_m
                 u_n^\ast(\bx,t) u_m(\bx,t)\nonumber\\
             & &\!\!+\sqrt{N_0}\ssum \left\{\alpha_n^\dagger u_n^\ast
                 (\bx,t) f(\bx,t)+\mbox{h.c.}\right\}\nonumber\\
             & &\!\!+N_0|f(\bx,t)|^2, 
\label{hatn}
\end{eqnarray} 
where 
\begin{eqnarray}
f(\bx,t)&=&u_0(\bx,t)\nonumber \\
&&-\ssum \gamma_n\{u_n(\bx,t)+i\varepsilon_n
         \int_0^t \! d\lambda\, u_n(\bx,\lambda)\}.
\end{eqnarray}
The thermal average of $\hat n(\bx,t)$ is given by 
\begin{equation}
n(\bx,t)=Z^{-1}\mbox{\bf Tr}\left[ e^{-\beta H_B}\hat n(\bx,t)\right],
\end{equation}
since the density matrix does not change in time in 
the Heisenberg picture.
Making use of the eqs.(\ref{HB1}) and (\ref{hatn}), finally we find 
\begin{eqnarray}
\lefteqn{n(\bx,t)=} \nonumber \\
&&N_0|u_0(\bx,t)-\ssum\gamma_n \{u_n(\bx,t)+i\varepsilon_n
            \int_0^t\! d\lambda\, u_n(\bx,\lambda)\}|^2\nonumber\\
        & &+\ssum\frac{1}{e^{\beta \varepsilon_n}-1} |u_n(\bx,t)|^2.
\label{n3}
\end{eqnarray}

As seen in the definition (\ref{defun}), the functions $u_n(\bx,t)$ represent 
the wave functions of free motion started from the harmonic 
oscillator eigenfunctions. Comparing the above $n(\bx,t)$ with 
$n(\bx)$ given in (\ref{n2}), we can see that $n(\bx,t)$ is the 
result of free evolution of $n(\bx)$, except for the term  
$i\varepsilon_n \int_0^t\! d\lambda\, u_n(\bx,\lambda)$. This term gives 
the effect of the interaction between particles. If we drop 
this term, the integration of $n(\bx,t)$ with respect to $\bx$
becomes the same as N in eq.(\ref{N2}), because of the orthogonality of
\{$u_n(\bx,t)$\}:
\begin{equation}
\int \! d \bx u_n^\ast (\bx,t) u_m(\bx,t) = \delta_{nm}.
\end{equation}
This means that, in our approximation, the time dependence of the fraction $f_0$
 stems from the interaction between particles.

In summary, the two main results of this work are as follows: (I) 
The ground 
state particle distribution in a Bose condensate 
under the trapping potential, seen from 
the first term in (\ref{n2}), is different from 
the noninteracting case, where the distribution is expected to be 
$|u_0(\bx)|^2$. (II) The time evolution of the particle distribution of 
a condensed Bose gas immediately after switching off the trapping 
potential suddenly can be calculated analytically as in (\ref{n3}).  
In (\ref{n2}) and (\ref{n3}) the parameter $\gamma_n$,
representing the interaction
effect, is crucial quantitatively. Simple manipulations of (\ref{gamman})
lead to
\begin{eqnarray}
\gamma_n&=& \frac{4\pi a}{(2\pi)^{3/2} \varepsilon_n}\, N_0 \prod_{i=1}^3
\omega_i^{1/2}\left(-\frac{1}{2}\right)^{n_i/2}
 \sqrt{\frac{(n_i-1)!!}{n_i!!}} \nonumber \\
&&\qquad\mbox{for all even $n_i$} \nonumber\\
& =& 0 \qquad \mbox{otherwise}
\end{eqnarray}
where $g=4\pi a$ and $a$ is the $s$-wave scattering length. 
To estimate the numerical order of the interaction effects
in our calculations, let us see the values of the parameters in
the experiment \cite{JILA} using Rb atoms.
There we have $N_0=2 \times 10^3$, $a=110 a_0$ 
($a_0$:Bohr radius), $\omega_3=\sqrt{8}\omega_1=\sqrt{8}\omega_2=200\pi\,$Hz,
then the leading $\gamma_n$ for $n=(2,0,0)$ or $(0,2,0)$ and the next-leading 
one for $n=(2,2,0)$ become -0.23 and 0.06, respectively.

Our calculations are based on Bogoliubov theory, adapted 
to a system trapped by a harmonic oscillator potential. 
Our approximation scheme is to take only the dominant terms 
in $N_0^{-1/2}$ expansion, but is still nonperturbative with respect 
to $g$ (the interaction between particles). Higher order calculations in 
$N_0^{-1/2}$ expansion is under study.  The future task is to combine this
field theoretical approach with nonequilibrium 
Thermo Field Dynamics \cite{Ari,Ume}.

This work was contributed to the International Workshop
on Thermal Field Theories and Their Applications 
in Regensburg, 10.-14. Aug. 1998.

\end{document}